\documentclass[conference]{IEEEtran}

\usepackage[a4paper, total={184mm,239mm}]{geometry}
\def\BibTeX{{\rm B\kern-.05em{\sc i\kern-.025em b}\kern-.08em
    T\kern-.1667em\lower.7ex\hbox{E}\kern-.125emX}}

\usepackage{booktabs}

\usepackage{url}
\usepackage{enumitem}
\usepackage{algorithmic}

\usepackage[linesnumbered]{algorithm2e}
\usepackage[utf8]{inputenc} 
\usepackage[T1]{fontenc}    
\usepackage{url}            
\usepackage{booktabs}       
\usepackage{amsfonts}       
\usepackage{nicefrac}       
\usepackage{microtype}      
\usepackage{xcolor}         
\usepackage{cite}
\usepackage{amsmath}
\usepackage{colortbl}
\usepackage{subcaption}
\usepackage{graphicx}
\usepackage{svg}
\usepackage{multirow}
\usepackage{wrapfig}
\usepackage{threeparttable}
\usepackage{tikz}
\usetikzlibrary{matrix}
\usepackage{makecell}
\usepackage{tablefootnote}

\usepackage{color}

\begin{document}
\title{BoolGebra: Attributed Graph-learning for Boolean Algebraic Manipulation}

\newcommand{\cy}[1]{\textcolor{red}{#1}}

\newcommand{\yj}[1]{\textcolor{green}{#1}}

\title{BoolGebra: Attributed Graph-learning for Boolean Algebraic Manipulation\\
\thanks{Identify applicable funding agency here. If none, delete this.}
}

\author{\IEEEauthorblockN{Yingjie Li$^1$, Anthony Agnesina$^2$, Yanqing Zhang$^3$, Haoxing Ren$^2$, Cunxi Yu$^1$
}
\IEEEauthorblockA{$^1$\textit{University of Maryland, College Park}, College Park, MD, USA \\
$^2$\textit{NVIDIA}, Austin, TX, USA \\
$^3$\textit{NVIDIA}, Santa Clara, CA, USA \\
\{yingjiel, cunxiyu\}@umd.edu, \{aagnesina, yanqingz, haoxingr\}@nvidia.com}}

\maketitle

\begin{abstract}

Boolean algebraic manipulation is at the core of logic synthesis in Electronic Design Automation (EDA) design flow. Existing methods struggle to fully exploit optimization opportunities, and often suffer from an explosive search space and limited scalability efficiency. 
This work presents BoolGebra, a novel attributed graph-learning approach for Boolean algebraic manipulation that aims to improve fundamental logic synthesis.
BoolGebra incorporates Graph Neural Networks (GNNs) and takes initial feature embeddings from both structural and functional information as inputs. A fully connected neural network is employed as the predictor for direct optimization result predictions, significantly reducing the search space and efficiently locating the optimization space. 
The experiments involve training the BoolGebra model w.r.t design-specific and cross-design inferences using the trained model, where BoolGebra demonstrates generalizability for cross-design inference and its potential to scale from small, simple training datasets to large, complex inference datasets. Finally, BoolGebra is integrated with existing synthesis tool ABC to perform end-to-end logic minimization evaluation w.r.t SOTA baselines.


\end{abstract}

\section{Introduction}
\label{sec:introduction}

Logic optimization is an essential stage in the design automation flow for digital systems as the performance of the system at logic level can have significant impacts on the final chip area, timing closure, and the power efficiency of the system~\cite{bjesse2004dag, haaswijk2018deep, brayton1990multilevel,amaru2015majority,mishchenko2007abc,yu2020flowtune,neto2022flowtune,yu2018developing}. Logic optimization is a technology-independent circuit optimization at the logic level, aiming to minimize the logic operations and logic level in the circuit while maintaining the original functionality of the circuit. The key methodologies of modern logic optimization techniques are conducted on multi-level technology-independent representations such as And-Inverter-Graphs (AIGs) \cite{mishchenko2006dag,mishchenko2005fraigs, yu2016dag,yu2017fast} and Majority-Inverter-Graphs (MIGs) \cite{amaru2015majority, soeken2017exact} of the digital logic, and XOR-rich representations for emerging technologies such as XOR-And-Graphs \cite{ccalik2019multiplicative} and XOR-Majority-Graphs \cite{haaswijk2017novel}. 
Existing state-of-the-art (SOTA) Directed-Acyclic-Graphs (DAGs) aware Boolean optimization algorithms, such as structural \texttt{rewriting}~\cite{mishchenko2006dag, riener2022boolean, haaswijk2018sat}, \texttt{resubstitution}~\cite{brayton2006scalable}, and \texttt{refactoring}~\cite{mishchenko2006dag} in ABC~\cite{mishchenko2007abc}, are conducted on the AIG data structure of the design following a domain-specific design concept, i.e., existing algorithms are implemented in a stand-alone fashion with single optimization operation in the single DAG-aware traversal.

\begin{figure*}
    \centering
    \begin{subfigure}[b]{0.19\textwidth}
    \centering
        \includegraphics[width=1\textwidth]{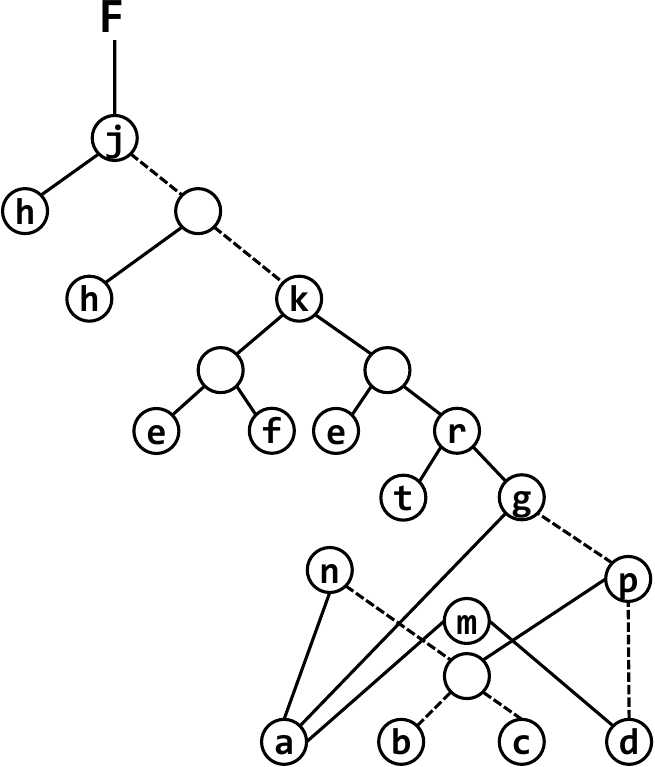}
        \caption{original AIG}
        \label{fig:aig_ori}
    \end{subfigure}
    \hfill
    \begin{subfigure}[b]{0.19\textwidth}
    \centering
        \includegraphics[width=1\textwidth]{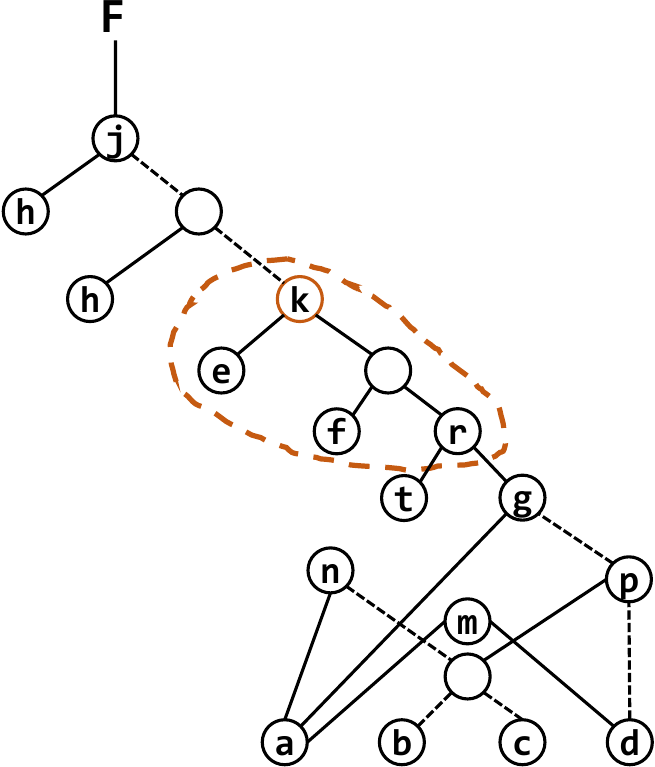}
        \caption{AIG with \texttt{rw}}
        \label{fig:aig_rw}
    \end{subfigure}
    \hfill
    \begin{subfigure}[b]{0.19\textwidth}
    \centering
        \includegraphics[width=1\textwidth]{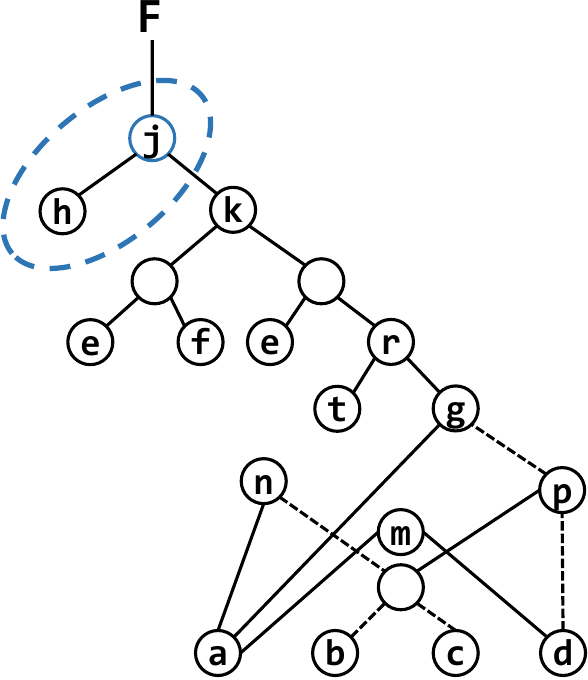}
        \caption{AIG with \texttt{rf}}
        \label{fig:aig_rf}
    \end{subfigure}
    \hfill
    \begin{subfigure}[b]{0.20\textwidth}
    \centering
        \includegraphics[width=1\textwidth]{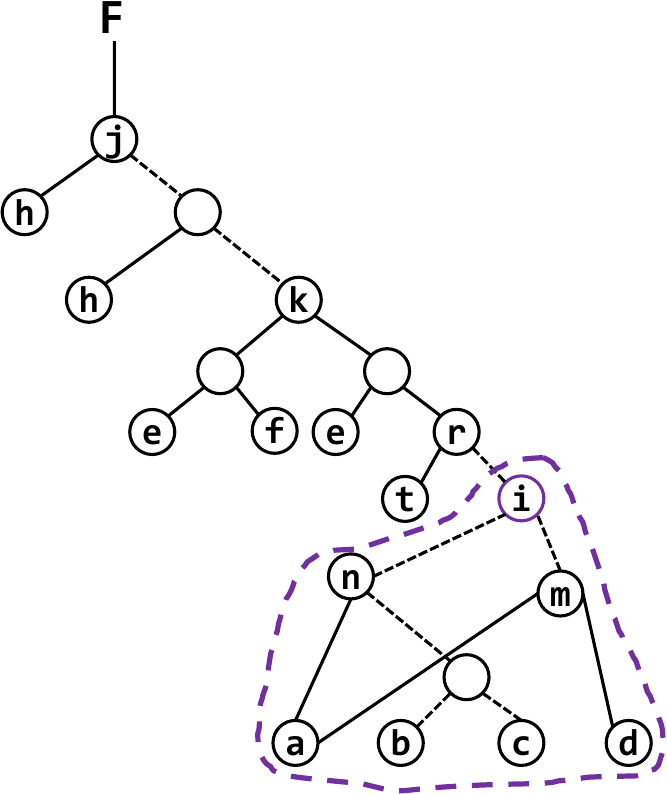}
        \caption{AIG with \texttt{rs}}
        \label{fig:aig_rs}
    \end{subfigure}
    \hfill
    \begin{subfigure}[b]{0.19\textwidth}
    \centering
\includegraphics[width=1\textwidth]{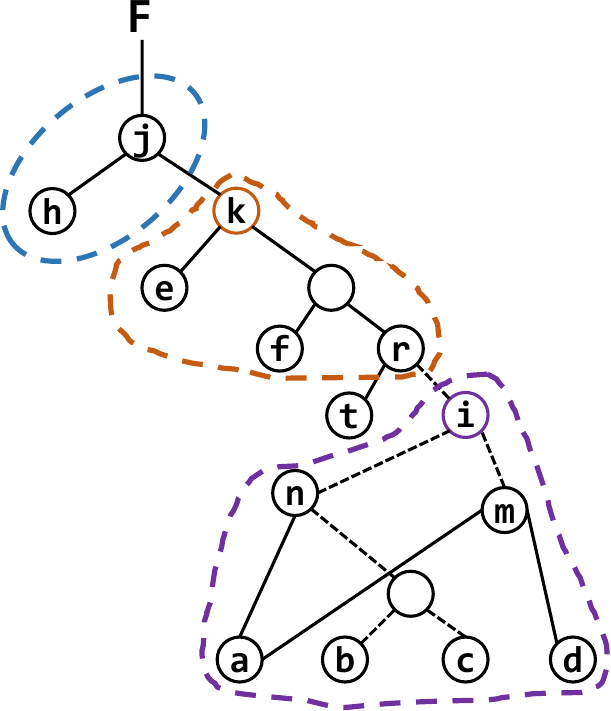}
        \caption{AIG with Algorithm \ref{alg:orch}}
        \label{fig:aig_orch}
    \end{subfigure}
    
    \caption{The optimized graph produced by stand-alone optimization operations and orchestration operation: (a) original AIG, graph size is 21; (b) optimized AIG with stand-alone \texttt{rw}, graph size is 19; (b) optimized AIG with stand-alone \texttt{rf}, graph size is 19; (c) optimized AIG with stand-alone \texttt{rs}, graph size is 20; (d) optimized AIG with proposed Algorithm \ref{alg:orch}, graph size is 16.
    }
    \label{fig:aig}
\end{figure*}

However, modern digital designs with increasing complexity encompassing millions of logic gates, engender an expansive exploration space. These designs pose substantial challenges to finding optimal solutions using SOTA standalone optimizations incorporated within extant logic synthesis tools. Two major complications emerge from such an approach. \textbf{Firstly}, a significant amount of potential optimizations are overlooked as the process contemplates only a single optimization operation during the traversal of the AIGs. This lack of holistic consideration could trap the optimization within suboptimal local minima. To illustrate, let us consider node $j$ in Figure \ref{fig:aig}, which is eligible for \texttt{refactor} operation. However, due to the solitary nature of the current optimization method, opportunities for \texttt{rewrite} optimization are disregarded. \textbf{Secondly}, the incorporation of multiple optimization techniques within a single AIG optimization exponentially enlarges the search space. In existing SOTA optimization, the optimization space is $O(N)$ for a design containing $N$ AIG nodes. However, when integrating multiple optimization techniques, such as \texttt{rewriting}, \texttt{resubstitution}, and \texttt{refactoring} within a single traversal, the optimization space escalates to $O(3^{N})$. This increase in complexity can pose a significant challenge in searching optimal solutions.


Our work proposes new solutions to AIG optimizations targeting the aforementioned two challenges. First, in existing logic optimizations on AIG, the operation is checked transformability and applied to the AIG node sequentially following the topological order, i.e., the optimization for each node is independent from each other, which provides the opportunity to explore different optimizations for nodes in the single AIG traversal. Thus, BoolGebra aims to discover novel manipulation over Boolean networks by integrating \texttt{rewriting}, \texttt{resubstitution}, and \texttt{refactoring} in single AIG traversal. 

Second, to deal with the exponentially increased search space with multiple optimizations, we see the potential for employing machine learning (ML) for electronic design automation (EDA) tasks~\cite{alrahis2021gnn,zhao2022graph,he2021graph,wu2022survey,wu2023gamora,yu2020decision,yin2023respect,yu2019painting}, as an aid to conventional design solutions. The circuit netlists or Boolean networks are represented as graphs, thus Graph Neural Networks (GNNs) are our natural choice for the logic optimization tasks based on AIGs. 
Specifically, to provide fast yet accurate optimization guidance for logic optimizations, we propose a graph-learning based neural network, which takes the design AIG as input and produces the optimized AIG result directly. Based on the fast-acquired inference result, the search space is thus shrunk, which aids the logic optimization to locate the optimized solution efficiently. Once the model is well trained, it shows its generalization capability from small, simple to large, complex designs.

The main contributions are summarized as follows:
\begin{itemize}[leftmargin=*]
    
    \item We develop a GNN-based predictor with function-aware embedding method for fast estimation of logic optimization performance w.r.t the AIG structure and per node algebraic manipulation assignment. 
    \item Exploit the capabilities of the GNN-based predictor, i.e., BoolGebra, a tool designed for rapid sampling and design space exploration. BoolGebra aims to locate unseen orchestrated Boolean manipulation solutions.
    \item We evaluate BoolGebra in a variety of optimization scenarios, encompassing both design-specific and cross-design optimization. Through these evaluations, BoolGebra demonstrates robust generalizability across multiple design environments.
    \item BoolGebra is integrated with the ABC synthesis framework \cite{mishchenko2007abc}, facilitating comprehensive end-to-end AIG reduction evaluation w.r.t ABC SOTAs.

\end{itemize}

\section{Preliminary}

\subsection{Boolean Networks and AIGs}
\label{sec:background_BN}
A Boolean Network is a directed acyclic graph (DAG) denoted as $G=(V, E)$ with nodes $V$ representing logic gates (Boolean functions) and edges $E$ representing the connections between gates. The input of a node is called its \textit{fanin}, and the output of the node is called its \textit{fanout}. The node $v \in V$ without incoming edges, i.e., no \textit{fanins}, is the \textit{primary input} (PI) to the graph, and the nodes without outgoing edges, i.e., no \textit{fanout}, are \textit{primary outputs} (PO) to the graph. The nodes with incoming edges implement Boolean functions. 

\textit{And-Inverter Graph} (AIG) is one of the typical types of DAGs used for logic manipulation, where the nodes in AIGs are all two-inputs AND gates, and the edges represent whether the inverters are implemented. An arbitrary Boolean Network can be transformed into an AIG by factoring the SOPs of the nodes, and the AND gates and OR gates in SOPs are converted to two-inputs AND gates and inverters with DeMorgan's rule. There are two primary metrics for evaluation of an AIG, i.e., \textit{size}, which is the number of nodes (AND gates) in the graph, and \textit{depth}, which is the number of nodes on the longest path from PI to PO (the largest level) in the graph. 

A \textit{cut} C of node $v$ includes a set of nodes of the network. The nodes included in the \textit{cut} of node $v$ are called \textit{leaves}, such that each path from a PI to node $v$ passes through at least one leaf. The node $v$ is called the \textit{root} of the \textit{cut} C. The cut size is the number of its leaves and the node itself. A cut is $K$-feasible if the number of nodes in the cut does not exceed $K$. 
The optimization of Boolean networks can be conducted with the AIGs efficiently~\cite{hosny2020drills, yu2018developing,li2023dag}. The AIG-based optimization process in the single traversal of the graph usually involves two steps: \textbf{(1)} \textit{transformability check} -- check the transformability w.r.t the optimization operation for the current node; \textbf{(2)} \textit{graph updates} {( \texttt{Dec\_GraphUpdateNetwork} in ABC \cite{mishchenko2007abc}) -- apply the applicable optimization operation at the node to realize the transformation and update the graph for the next unseen node.

\subsection{DAG-Aware Logic Synthesis}
\label{sec:background_DAG_syn}

DAG-aware logic synthesis approaches leverage Boolean algebra at direct-acycle-graph (DAG) representations to reduce the logic complexity and size, while preserving the original functionality of the circuit, aiming to the enhanced performance of area, delay, power, etc., in the final digital systems. 
This process involves a variety of optimization techniques and algorithms, e.g., node rewriting, structural hashing, and refactoring. In this work, we are particularly interested in exploring DAG-aware logic synthesis on AIGs representations.  
 
\noindent
\textbf{Rewriting}, noted as \texttt{rw}, is a fast greedy algorithm for optimizing the graph size. It iteratively selects the AIG subgraph with the current node as the root node and replaces the selected subgraph with the same functional pre-computed subgraph with a smaller size to realize the graph size reduction. Specifically, it finds the 4-feasible cuts as subgraphs for the node while preserves the number of logic levels~\cite{mishchenko2006dag}. 
For example, Figure \ref{fig:aig_rw} shows the optimization of the original graph in Figure \ref{fig:aig_ori} with \texttt{rw}. The algorithm visits each node in the topological order and checks the transformability of its cut w.r.t \texttt{rw}. It will skip the node and visit the next one if not applicable. In Figure \ref{fig:aig_rw}, the node $j$ is skipped for \texttt{rw}, and node $k=efr$ is optimized with \texttt{rw}, resulting in the node reduction of $2$ for the AIG optimization.

\noindent
\textbf{Refactoring}, noted as \texttt{rf}, is a variation of the AIG \texttt{rewriting} using a heuristic algorithm~\cite{brayton2006scalable} to produce a large cut for each AIG node. Refactoring optimizes AIGs by replacing the current AIG structure with a factored form of the cut function. It can also optimize the AIGs with the graph depth.
For example, Figure \ref{fig:aig_rf} shows the optimization of the original AIG with \texttt{rf}. The node $j=h(\overline{h} + k)$ is optimized with \texttt{rf} based on DeMorgan’s rule, where $j=hk$, and similarly for the node $g$. As a result, the optimized graph with \texttt{rf} has a graph size of $17$ with $3$ nodes reduction and $2$ depth reduction.

\noindent
\textbf{Resubstitution}, noted as \texttt{rs}, optimizes the AIG by replacing the function of the node with the other existing nodes (\texttt{divisors}) already present in the graph, which is expected to remove the redundant node in expressing the function of the current node. For example, in Figure \ref{fig:aig_ori}, the node $g=a\overline{p}$, $p=\overline{b}\overline{c}\overline{d}$, ${n}={(b + c)}{a}$, $m=ad$, i.e., $g={{m} + {n}} = \overline{\overline{m}\cdot\overline{n}}$, which can be resubstituted with node $i=\overline{m}\cdot\overline{n}$, and node $p$ is removed from the graph. As a result, with \texttt{rs}, the original AIG is optimized in graph size by $1$ node reduction.

\subsection{Graph Neural Networks in EDA}

Graph learning has emerged as a powerful method for understanding complex relationships in various domains, including social networks, biological systems, and natural language processing. In the context of hardware design, graph learning can be employed to model and analyze the intricate connections among design components, such as gates, wires, and registers, to optimize design performance and effectively analyze design characteristics. For example, dataflow graphs, circuit netlist, and Boolean networks (BNs) can be easily represented as graphs, making graph neural networks (GNNs) a natural fit for classifying sub-circuit functionality from gate-level netlists~\cite{alrahis2021gnn}, analyzing the impacts of circuit rewriting~\cite{zhao2022graph}, predicting arithmetic block boundaries~\cite{he2021graph, wang2022functionality}, improving the fidelity of early-stage design exploration \cite{ustun2020accurate,wu2021ironman,lu2022placement,ustun2019lamda,pal2022machine}, hardware security \cite{alrahis2023graph,yu2021hw2vec}, technology mapping \cite{neto2021read,neto2021slap}, etc. However, we believe there does not exist works in discovering novel Boolean manipulation techniques in fundamental logic synthesis.


\begin{figure}[!htb]
    \centering
    \begin{subfigure}[b]{0.24\textwidth}
    \centering
        \includegraphics[width=1\textwidth]{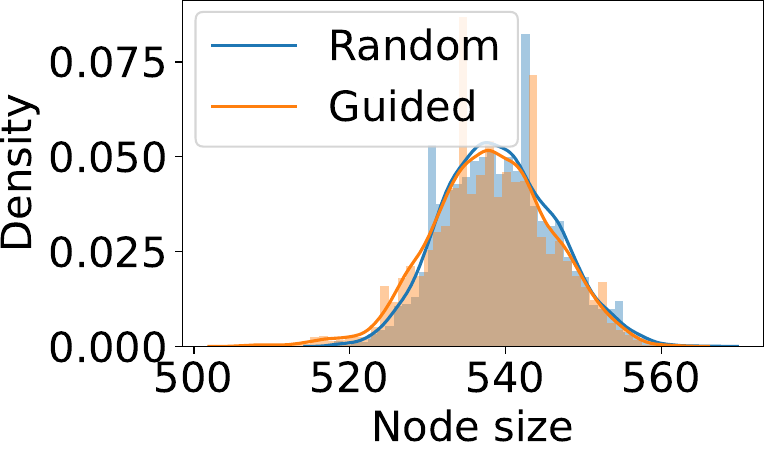}
        \caption{b11}
        \label{fig:sample_b11}
    \end{subfigure}
    \hfill
    \begin{subfigure}[b]{0.24\textwidth}
    \centering
        \includegraphics[width=1\textwidth]{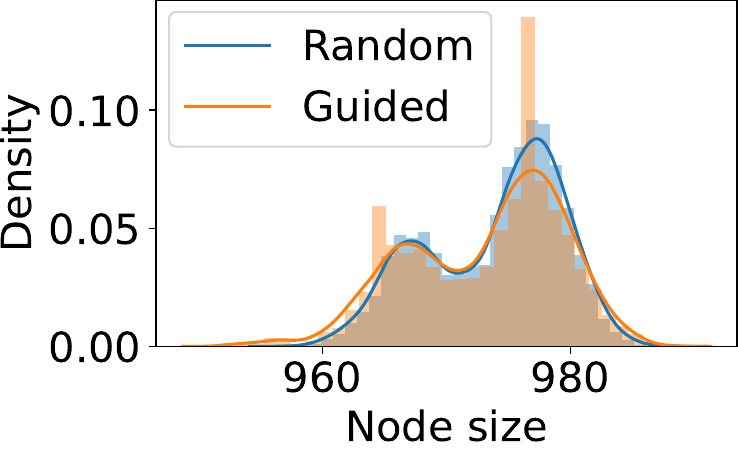}
        \caption{b12}
        \label{fig:sample_b12}
    \end{subfigure}
    \hfill
    \begin{subfigure}[b]{0.24\textwidth}
    \centering
        \includegraphics[width=1\textwidth]{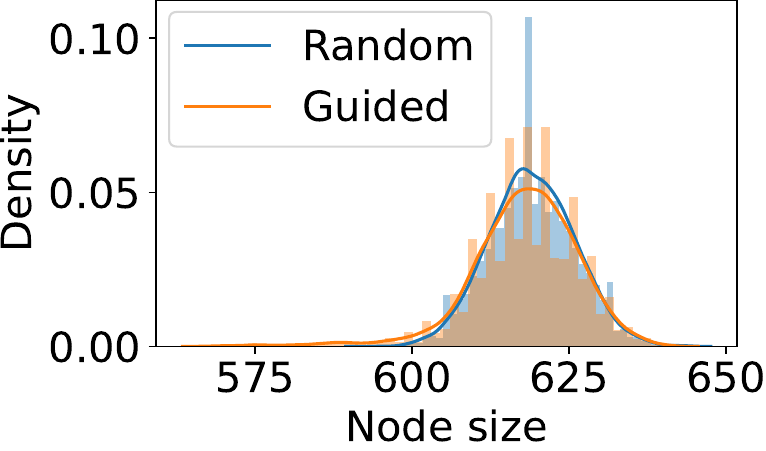}
        \caption{c2670}
        \label{fig:sample_c2670}
    \end{subfigure}
    \hfill
    \begin{subfigure}[b]{0.24\textwidth}
    \centering
        \includegraphics[width=1\textwidth]{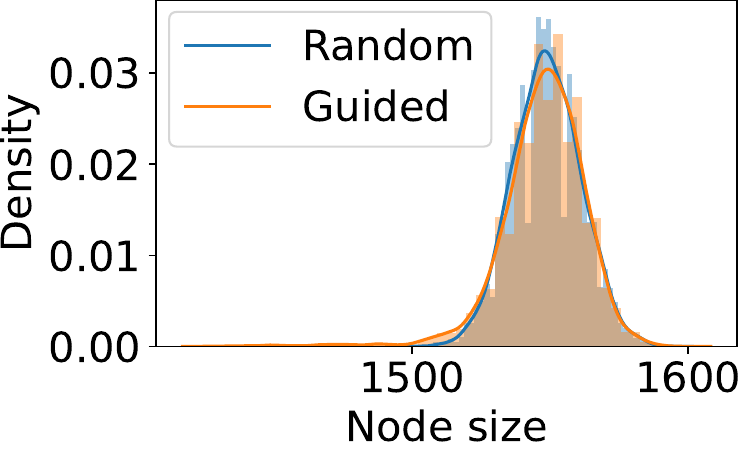}
        \caption{c5315}
        \label{fig:sample_c5315}
    \end{subfigure}
    
    \caption{The optimization quality distribution with 6000 samples of purely random sampling and priority guided sampling. 
    }
    \label{fig:sample}
\end{figure}

\section{Approach}
\label{sec:approach}


In this section, we first provide the studies on the optimization opportunities in the single traversal of AIG for the orchestration optimization (Section \ref{sec:space}). Then we introduce the method with orchestrating multiple optimizations with \texttt{rw}, \texttt{rs}, and \texttt{rf} by taking arbitrary assigned optimization at each node (Section \ref{sec:alg_orch}). We further introduce the BoolGebra flow, including the node feature embeddings and the GNN model for logic optimization prediction (Section \ref{sec: predict_model}), to shrink the orchestrated optimization search space and locate the optimization samples efficiently. 

{ 
\subsection{Optimization Space and Motivation}
\label{sec:space}


Our analysis (Figure \ref{fig:sample}) aims to explore and comprehend the optimization opportunities and space associated with minimizing AIGs via Boolean manipulations. Specifically, we employ a random sampling approach to generate various manipulation decisions for each node in the AIG, including \texttt{rw}, \texttt{rs}, and \texttt{rf}. Each random sample represents a potential Boolean synthesis solution, resulting in an updated AIG. Note that these samples adhere to the principles of Boolean algebra, ensuring the functional equivalence of all generated graphs.


Figure \ref{fig:sample} illustrates the results obtained from randomly sampling 6,000 solutions for four different designs (b11, b12, c2670, and c5135) sourced from well-established ITC/ISCAS99 and ISCAS85 benchmarks. 
Our analysis reveals two key findings. Firstly, the choice of manipulation decisions for the Boolean network graphs significantly impacts the final quality-of-results in logic minimization. Secondly, the distribution of quality-of-results within the randomly sampled design space approximately follows a Gaussian distribution. Consequently, employing a random sampling method for the purpose of logic minimization, specifically in identifying solutions that offer the minimum AIG size, proves to be challenging or infeasible. This study affirms the presence of optimization opportunities and challenges, thereby motivating our work in leveraging graph learning techniques to identify superior Boolean manipulation methods. In Section \ref{sec:approach}, we will further discuss Guided sampling that optimizes our training performance.


\subsection{{Design Augmentation and Sampling}}
\label{sec:alg_orch}

Here, we introduce our implementation for generating such random sampling solution and evaluate its AIG reduction performance, which is illustrated in Algorithm \ref{alg:orch}. 

First, the algorithm takes the design in AIG representation $G(V, E)$ and the random decision for each node in vector $D$, which is stored in the file in csv format as inputs, where $D$ is the same size of $V$. The three optimizations are encoded in integer indices, i.e., $0$ for \texttt{rw}, $1$ for \texttt{rs}, and $2$ for \texttt{rf}. For example, $D[v] = 0$ indicates that for node $v$, the optimization $0$, i.e., \texttt{rw}, is assigned for optimization. Subsequently, following the topological order, the algorithm checks the transformability of each node with respect to the assigned optimization operation $D[v]$ (line 2). If the optimization is applicable to the node for logic optimization, it is executed, resulting in the subgraph (the cut for node $v$) being updated (lines 3, 7) and the node itself is excluded from subsequent iterations. Conversely, if the optimization cannot be applied to the node, it is skipped (line 5) for further optimizations. Algorithm \ref{alg:orch} is implemented in ABC\cite{mishchenko2007abc}.
}

\begin{algorithm}
\caption{Boolean manipulation sampling on AIGs}  
\label{alg:orch}
\footnotesize
   \SetKwInOut{KwIn}{Input}
     \SetKwInOut{KwIn}{Input}
    \SetKwInOut{KwOut}{Output}
    \KwIn{~$G(V,E) \leftarrow$ Boolean Networks/Circuits in AIG} 
    \KwIn{Per node manipulation decision $D$, $|D| = |V|$}
    \tcp{$D$ saved as an Array $D$, e.g., $D[1] = 0/1/2$ indicates the optimization for node $1$ is $0/1/2$, i.e., \texttt{rw}/\texttt{rs}/\texttt{rf}.}
    \KwOut{~Post-optimized AIG $G(V,E)$}
 \For{$v \in V$ in topological order}{
  \uIf{v \text{is transformable w.r.t} $D[v]$}{
        $ G(V,E) \xleftarrow{D[v]} G(V,E)$
        continue
     \tcp*[f]{check the transformability with the assigned optimization}
         }

  \Else{
    continue 
  }
  Update $G(V,E) \leftarrow$ \texttt{Dec\_GraphUpdateNetwork}, and exclude $v$ and transformed nodes from $V$
  }
\end{algorithm}

\begin{figure*}[!htbp]
    \centering
    \includegraphics[width=1\textwidth]{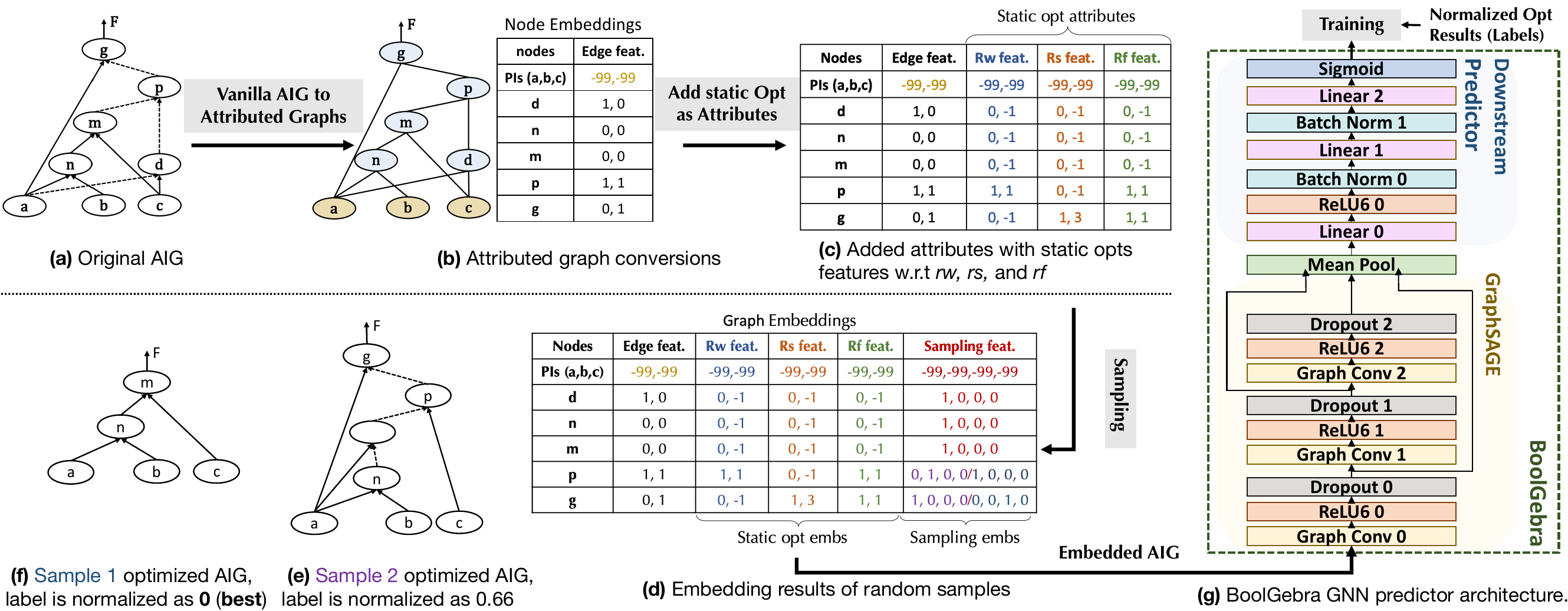}
    \caption{The design flow of BoolGebra including the feature embeddings with static and dynamic sampling embeddings ((a) -- (f)), and the model construction ((g)).}
    \vspace{-3mm}
    \label{fig:flow}
\end{figure*}

\subsection{{GNN-based prediction model}} \label{sec: predict_model}

With multiple optimizations orchestrated in the algorithm, the optimization space become even larger, resulting in difficulties in locating the optimal solution efficiently. For example, for the design \texttt{voter}, whose original size is $13758$, when orchestrating three operations in the logic optimization, the search space explodes to $3^{13758}$, which is hardly possible to do exact explorations. Thus, to shrink the search space and locate the optimization direction efficiently, a GNN-based prediction model is proposed with AIG structure and initial feature embeddings as inputs and predicts the optimization performance directly. Note that the GNN-based model has strong generalization capability, thus, the model can train on small datasets but apply to large datasets for efficient inference optimization results. 

In this section, we first introduce the dataset preparation for the model including the initial feature embeddings with structural and functional information for the AIG nodes, and the data normalization. Then, we introduce the structure of GNN-based model for logic optimization predictions.

\subsubsection{Dataset Preparation}

Utilizing the principles of message propagation and neighborhood aggregation within GNNs, we generate node embeddings for AIGs that effectively combine both structural and functional information.
The structural information, i.e., the edgelist for the AIG, is integrated by transmitting node embeddings along edges.
Second, the Boolean functional information can be encoded in node features.
The initial node features are in two categories: (1) static features, where the features are static w.r.t the AIG structure, and will not alter with different optimization assignments; (2) dynamic features, where the features change as different assignments are given, which introduce different optimization results. 

\noindent
\textbf{Feature Embedding} -- 
First, we extract the static features as shown in Figure \ref{fig:flow} for each node. For the PI nodes (node $a, b, c$ in Figure \ref{fig:flow}(a)), which have no fanins, the type features are all set as $-99$. For the nodes with fanins, we attach 4 sets of features with 2 bits for each set, including:

\textbf{1)-(a)} -- \textit{the characteristics of AIG edges}. In AIG, the nodes are all 2-input AND nodes with edges inverted or not. From the left input edge to the right input edge, if the input edge is inverted, it will be encoded as $1$; otherwise, encoded as $0$. {\textbf{Example}: As shown in Figure \ref{fig:flow}(b), the node $d$, whose left fanin edge is inverted and the right edge is not, is embedded with the node feature of $[1, 0]$}. Note that since AIG is an universal representation and edge weights represent complement or not, this embedding is capable to incorporate local Boolean function. 

\textbf{1)-(b)} -- \textit{\texttt{rw} transformability and local optimization gain} (third and fourth columns in Figure \ref{fig:flow}(c)). If \texttt{rw} is applicable for optimization at the node, then the third bit is set as $1$ and the fourth bit is set as its corresponding optimization gain, i.e., the number of AIG minimization with the optimization applied; otherwise, if \texttt{rw} is not appliable, the third bit is set as $0$, and the fourth bit is set as $-1$. {\textbf{Example}: In the AIG in Figure \ref{fig:flow}(a), for node $p$, it can be optimized with \texttt{rw} and its gain is $1$. Thus, the feature embedding for node $p$ is $[1, 1]$. However, for node $d$, which cannot be optimized by \texttt{rw}, its embedding is set as $[0, -1]$ at third and fourth bits.}; 

\textbf{1)-(c)} -- \textit{\texttt{rs} transformability and local optimization gain} (fifth and sixth columns in Figure \ref{fig:flow}(c)). {\textbf{Example}: In the AIG in Figure \ref{fig:flow}(a), the node $g$ can be optimized with \texttt{rs} and its gain is $3$. Thus, the fifth bit and sixth bit are set as $1$ and $3$, respectively. For other inapplicable nodes with \texttt{rs} such as node $p$, $m$, and $n$, the fifth bit is set as $0$ and the sixth bit is set as $-1$.}; 

\textbf{1)-(d)} -- \textit{\texttt{rf} transformability and local optimization gain} (seventh and eighth columns in Figure \ref{fig:flow}(c)). {\textbf{Example}: In Figure \ref{fig:flow}(a), the node $g$ can be optimized with \texttt{rf} and its gain is $1$, thus, its seventh and eighth bit in the feature embedding are set as $1$ and $1$, respectively. While for the node such as $m$, which is not applicable with \texttt{rf}, the seventh and eighth bits are $0$ and $-1$}. 

As a result, for each AIG node, its static feature embedding is an 8 dimensional vector containing information from its fanins, its transformability w.r.t orchestrated optimizations, i.e., \texttt{rw}, \texttt{rs}, and \texttt{rf}. {For example, in Figure \ref{fig:flow}, the node $g$ is embedded with features $[0, 1, 0, -1, 1, 3, 1, 1]$ for static feature embeddings.}. Note that these features are independent from the optimization sampling, which means they are static features dependent on the design structure only. 

Second, for the dynamic feature as shown in Figure \ref{fig:flow}(d), for each node, it is a 4-bit one-hot representation. The embedding for PIs is set as $[-99, -99, -99, -99]$ to indicate they are PIs with no fanins. Then, for logic nodes, the dynamic feature indicates which operation is practically applied to the node under the specific optimization sampling. The 4 bits indicate none of optimizations is applied, \texttt{rw} is applied, \texttt{rs} is applied, \texttt{rf} is applied, respectively. The practically applied operation is set as $1$ in one-hot representation while others are set as $0$. {For example, in Sample 2 (Figure \ref{fig:flow}(e)), node $p$ is optimized with \texttt{rw} and the dynamic embedding for node $p$ is thus $[0, 1, 0, 0]$ and node $g$ is not optimized and its dynamic embedding is $[1, 0, 0, 0]$.}

Note that the dynamic features vary according to the assignments under the current sampling, i.e., it depends not only on the design structure but also depends on the given optimization assignments. For example as shown in Figure \ref{fig:flow}, Figure \ref{fig:flow}(e) and Figure \ref{fig:flow}(f) are two different samples with two different dynamic assignments to node $p$ and $g$, which lead to different samples with different feature embeddings and the corresponding optimization labels, which is referred as ground truth labels in the model training.

\noindent
\textbf{Data Normalization} -- The data distribution has significant importance on the model training and its performance. In this work, we find several challenges w.r.t the datasets during the training process: (1) the model overfits easily as the direct optimization results vary in a small range compared to its size number. For example, for the design \texttt{b12}, whose original AIG size is $1002$, the best AIG optimization we observed is $952$ and the worst is $1002$. Thus the best AIG minimization is $50$, which is only 5\% of its size number $1002$. That is, the labels for the samples are ranging from 95\% to 100\% w.r.t its original size; (2) the training data can be less distinctive within limited random samples as shown in Figure \ref{fig:sample}, i.e., purely random sampling can result in less various data samples, which can even worsen the overfitting problem in the model training.

Targeting the two challenges, we first normalize the AIG minimization w.r.t the highest node reduction number in the optimization samples, i.e., we set ratio of the gap between the node reduction in the current sample to the most node reduction of all the samples in the dataset as the label for the current sample. For example, for the two samples shown in Figure \ref{fig:flow}(e) and Figure \ref{fig:flow}(f), the best optimization for the dataset with these two samples reduces the AIG for $3$ nodes as shown in Figure \ref{fig:flow}(f), thus, the label for Sample 1 (Figure \ref{fig:flow}(f)) is $0$. While for Sample 2 in Figure \ref{fig:flow}(e), the reduction node number is $1$ and its label is $0.66$. Thus, we normalize the node reduction for the optimization samples into the percentage w.r.t the best AIG minimization ranging between $[0, 1]$. In this case, the model aims to relatively identify the best optimization candidates, rather than predicting the optimization results directly. By acquiring the ratio number from the model, the selected most promising ''optimal'' samples with smaller ratio will be evaluated in ABC \cite{mishchenko2007abc} for final evaluation.

{To address the second challenge, we prioritize optimization techniques with minimal structural transformations during data sampling. Based on \cite{yu2020flowtune}, we prioritize \texttt{rw} over \texttt{rf} to minimize structural changes. We assign different priorities to sampling options (\texttt{rw}, \texttt{rs}, and \texttt{rf}) and apply the highest priority operation to each node when applicable. Other applicable optimizations are randomly applied to nodes with lower priority. Using the initial sample, we generate additional data samples through partial random assignments, sampling a percentage of AIG nodes (10\% to 90\%) to create training data. The priority-guided random sampling produces more data samples with improved performance, enhancing the modeling training with efficient optimization structures as shown in Figure \ref{fig:sample}.}


\subsubsection{GNNs Predictor}

As shown in Figure \ref{fig:flow}(g), the model takes the feature embedding at each node and the edgelist of the AIG as inputs and produces the AIG reduction as the prediction results. The model consists of two parts: the graph embedding part utilizing GraphSAGE \cite{hamilton2017inductive} for attributed graph feature extraction and aggregation, and the prediction part employing multi-layer fully connected neural networks for predictions.

In the first part, the feature and structure embeddings of the AIG are fed into the model, and three GraphSAGE Convolution (Conv) layers are utilized. Each Conv layer is accompanied by a dropout layer with a dropout rate of $0.1$ and a nonlinear layer (ReLU6). The hidden dimension of each Conv layer is $512$, and the output dimension is $64$. Following the feature embeddings obtained through GraphSAGE, the features are fused with a Pooling layer for concatenation with fully connected neural networks. The downstream model consists of three dense layers with output dimensions of $1000$, $200$, and $1$ respectively, in order to complete the regression downstream model. The first linear layer is connected to an additional nonlinear layer (ReLU6), followed by a Batch Norm layer (BatchNorm 0). The results from the Batch Norm layer are then connected to the second linear layer, which is combined with another Batch Norm layer (BatchNorm 1). Finally, the last linear layer is connected to a sigmoid layer to produce prediction results in the range of $[0, 1]$.

{
 
\subsection{BoolGebra Flow}

Finally, the aforementioned technical steps are integrated to form the BoolGebra flow. The BoolGebra flow comprises three key steps: \textbf{(1)} Random sampling of a large batch of Boolean manipulation decisions on a given AIG.
\textbf{(2)} Pruning the randomly sampled design space using the GNN-based predictor. \textbf{(3)} Evaluation of the top solutions based on the prediction results obtained in step 2 to derive the final AIG reduction outcomes.

 In this paper, we conduct our experimental results (Section \ref{sec:experiment}) using a single BoolGebra flow, where the sampling, pruning, and evaluation steps are performed only once. Specifically, we employ a sampling size of 600 per design, and the top 10 results from the prediction phase are subsequently evaluated to determine their effectiveness.

}

\begin{figure}[t]
    \centering
    \includegraphics[width=1\linewidth]{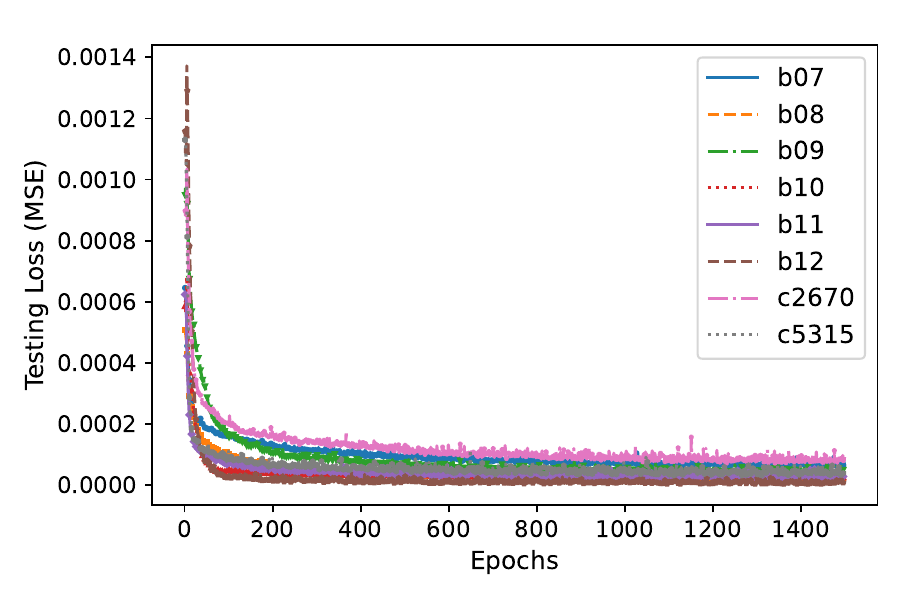}
    \caption{Design-specific testing loss w.r.t training epochs.}
    \label{fig:training_loss}
\end{figure}

\section{Experiments}
\label{sec:experiment}

\begin{figure*}[!htb]
  \centering

  \begin{subfigure}[b]{0.32\textwidth}
    \includegraphics[width=\textwidth]{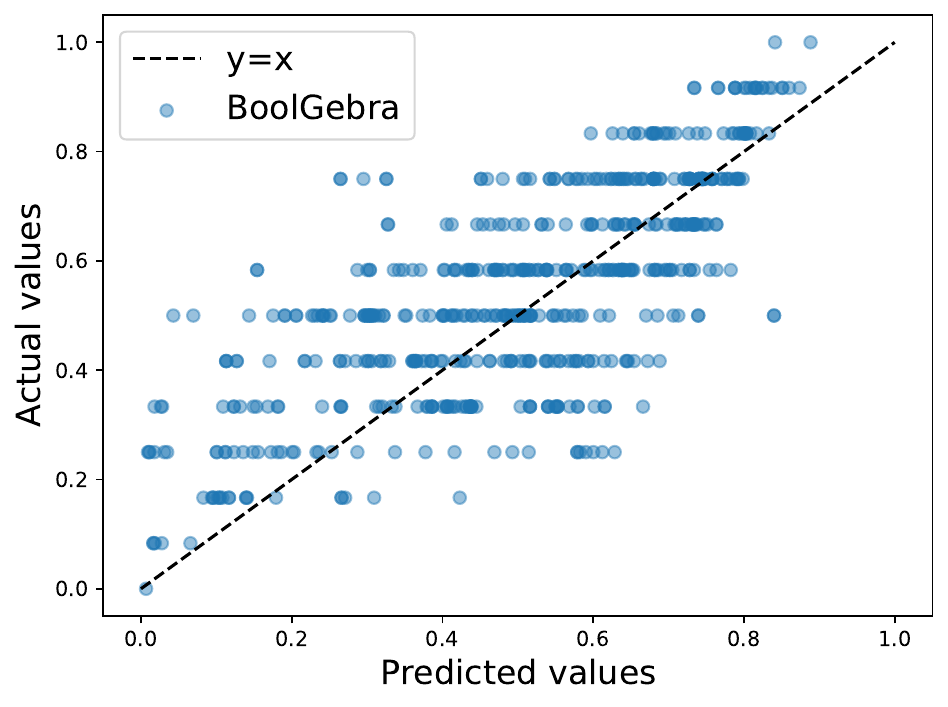}
    \caption{b07}
    \label{fig:spec_b07}
  \end{subfigure}
  \hfill
  \begin{subfigure}[b]{0.32\textwidth}
    \includegraphics[width=\textwidth]{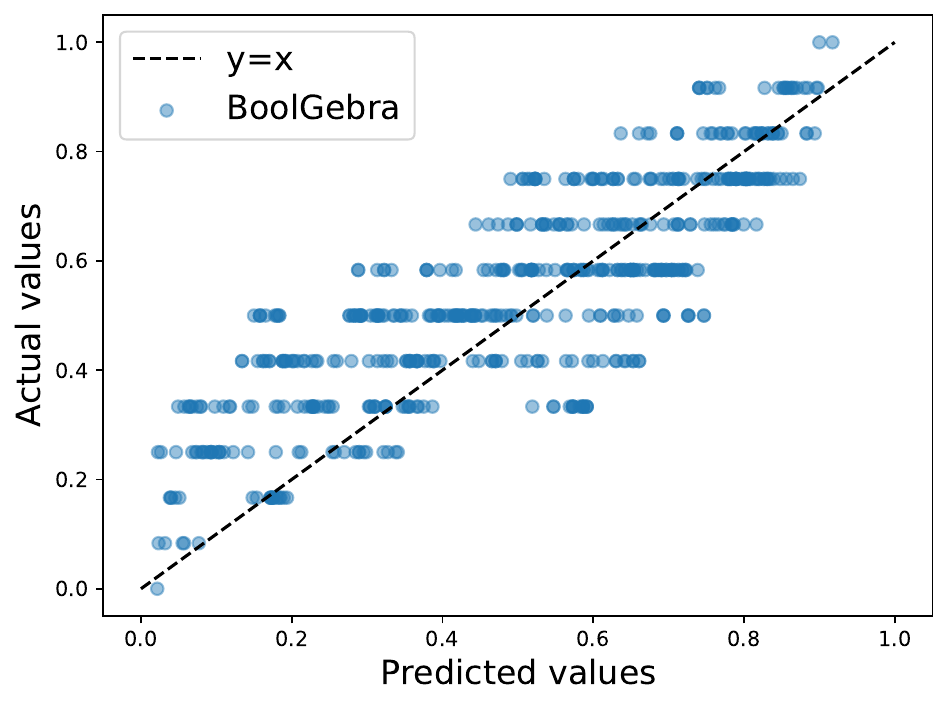}
    \caption{b10}
    \label{fig:spec_b10}
  \end{subfigure}
  \hfill
  \begin{subfigure}[b]{0.32\textwidth}
    \includegraphics[width=\textwidth]{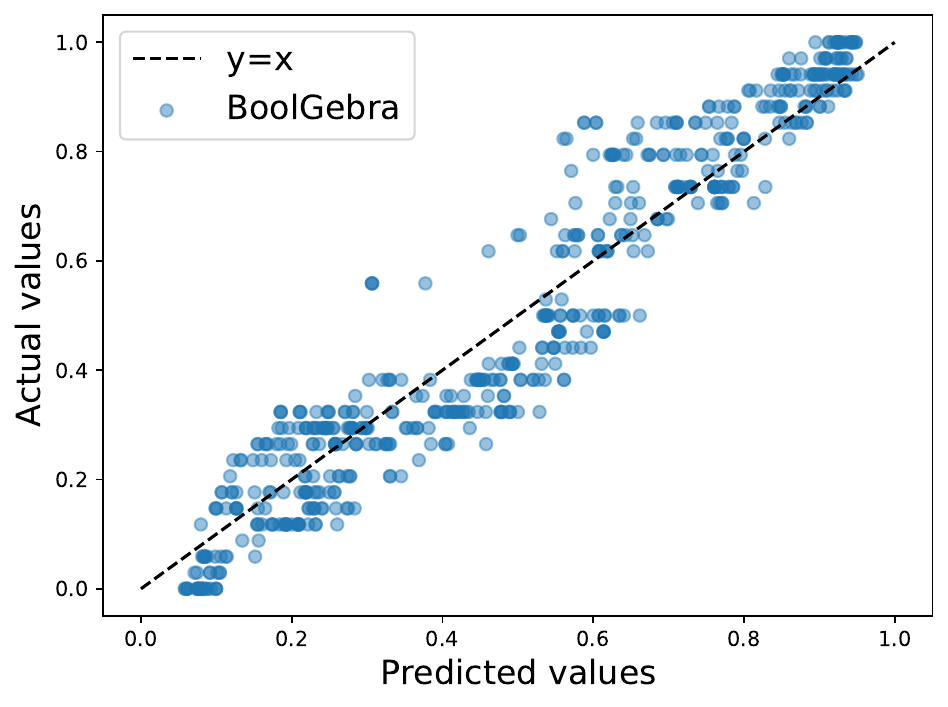}
   \caption{b12}
    \label{fig:spec_b12}
  \end{subfigure}
  \vspace{10pt}

  \begin{subfigure}[b]{0.32\textwidth}
    \includegraphics[width=\textwidth]{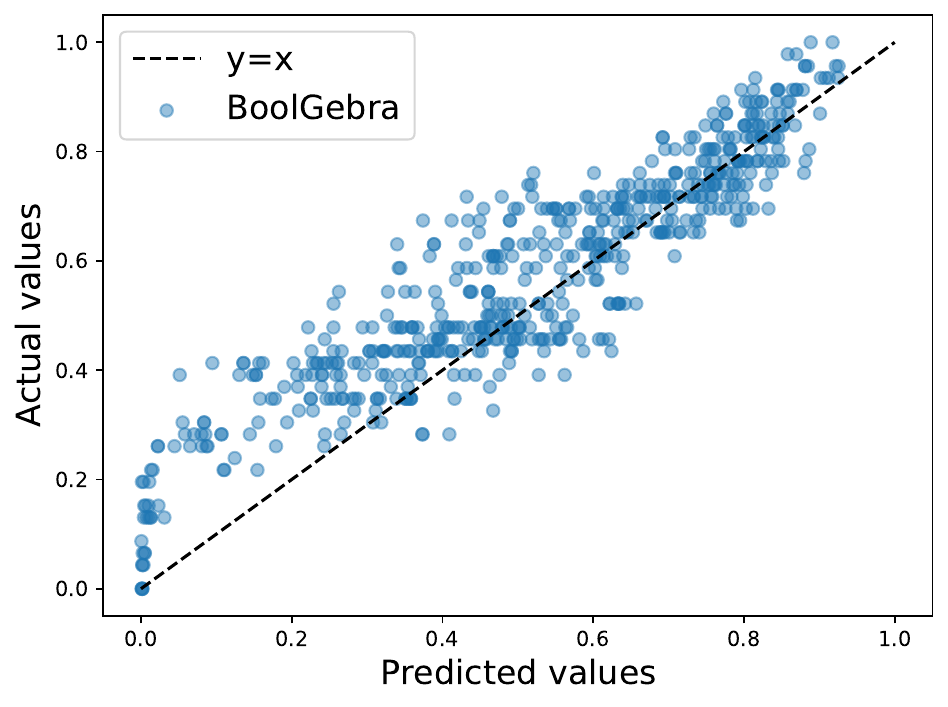}
    \caption{b11}
    \label{fig:spec_b11}
  \end{subfigure}
  \hfill
  \begin{subfigure}[b]{0.32\textwidth}
    \includegraphics[width=\textwidth]{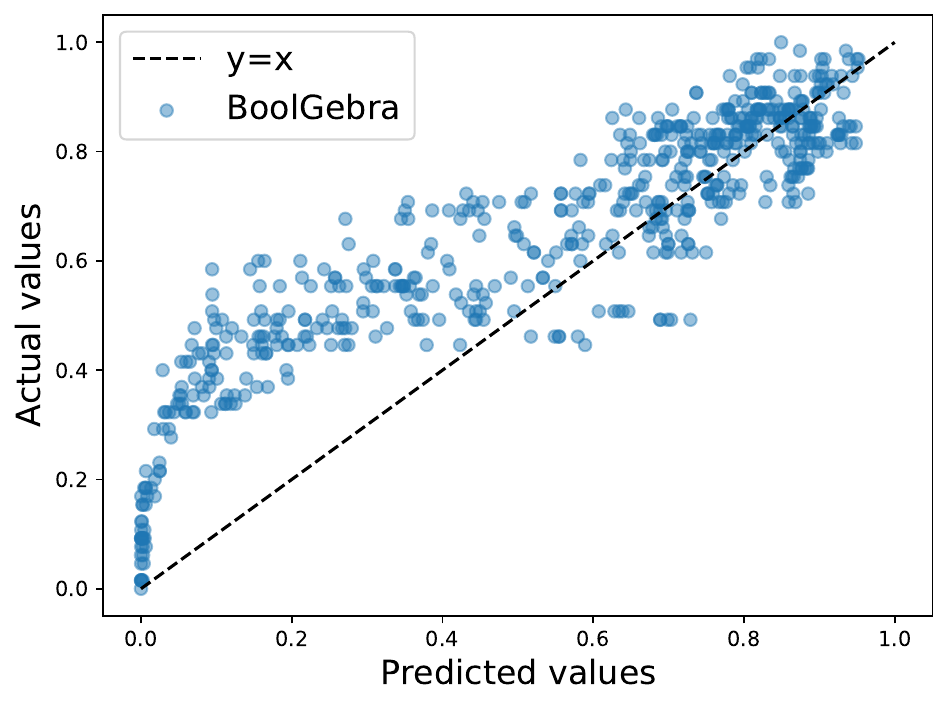}
    \caption{c2670}
    \label{fig:spec_c2670}
  \end{subfigure}
  \hfill
  \begin{subfigure}[b]{0.32\textwidth}
    \includegraphics[width=\textwidth]{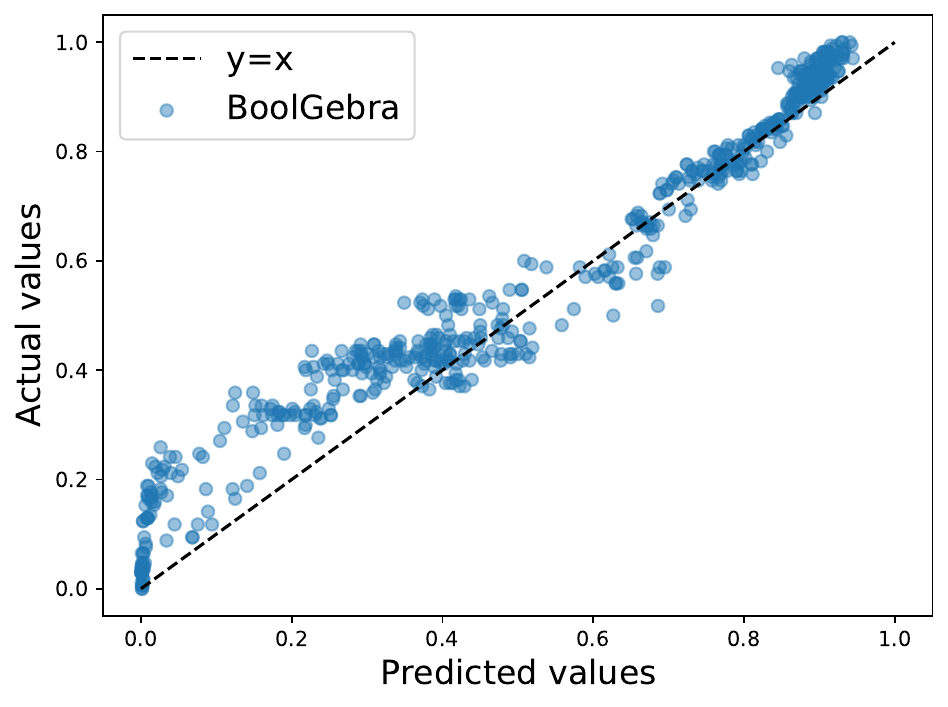}
    \caption{c5315}
    \label{fig:spec_c5315}
  \end{subfigure}

  \caption{Design-specific inference evaluation for predicting AIG minimization performance for given manipulation decisions. 
  Each sub-figure corresponds to an individual design, where the inference input are unseen randomly sampled decisions, where $x$-axis represents the normalized prediction and $y$-axis represents the normalized ground truth. Note that value "0" refers to the best quality-of-results and "1" refers to the worst.}
  \label{fig:specific_eval}
  \vspace{-3mm}
\end{figure*}

\begin{figure*}[!htbp]
\centering


\begin{subfigure}{0.32\textwidth}
  \centering
  \includegraphics[width=\linewidth]{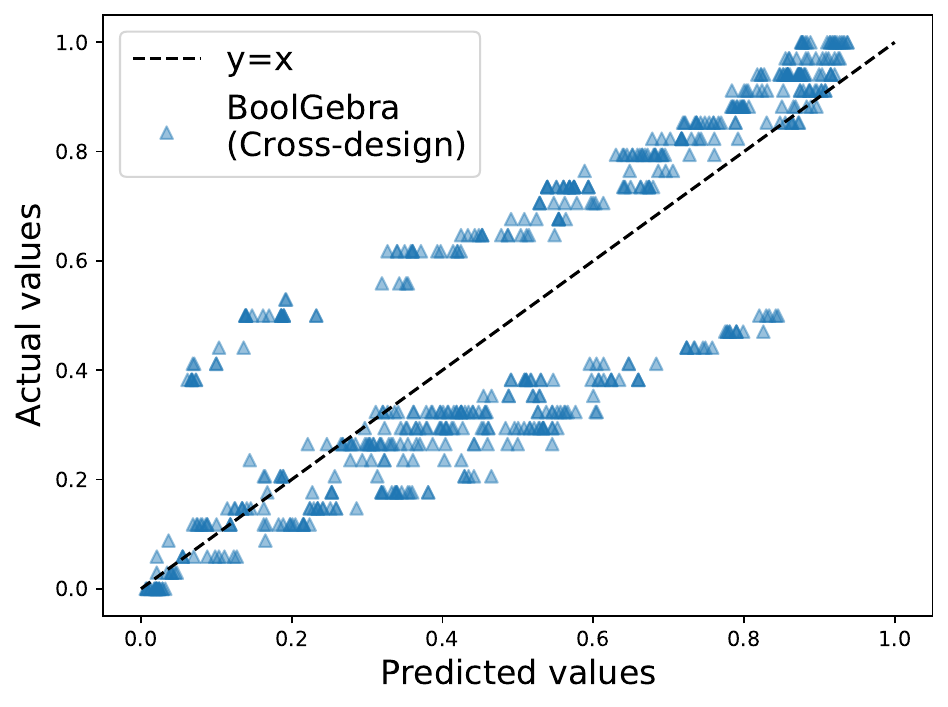}
  \caption{Testing: b12, Training: b11}
  \label{fig:b12_b11}
\end{subfigure}%
\hfill
\begin{subfigure}{0.32\textwidth}
  \centering
  \includegraphics[width=\linewidth]{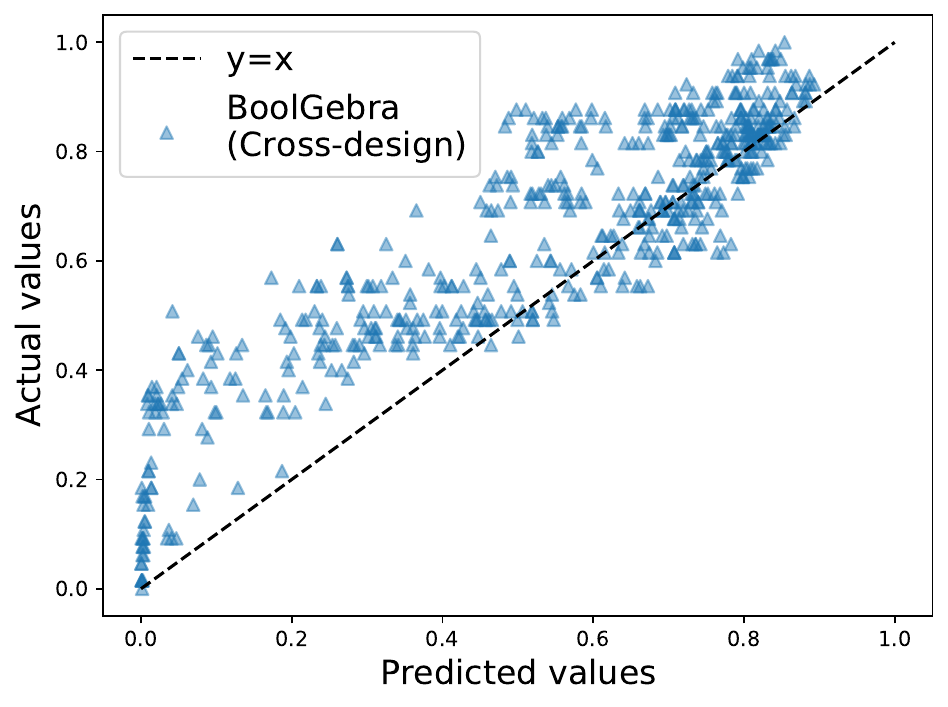}
  \caption{Testing: c2670, Training: b11}
  \label{fig:c2670_b11}
\end{subfigure}%
\hfill
\begin{subfigure}{0.32\textwidth}
  \centering
  \includegraphics[width=\linewidth]{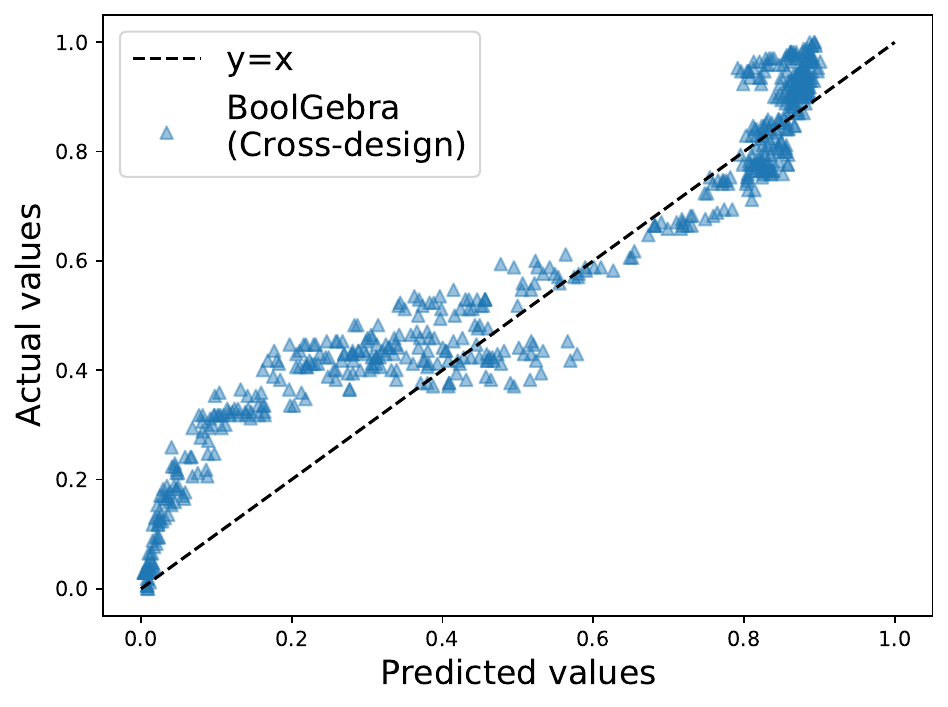}
  \caption{Testing: c5315, Training: b11}
  \label{fig:c5315_b11}
\end{subfigure}
\hfill

\begin{subfigure}{0.32\textwidth}
  \centering
  \includegraphics[width=\linewidth]{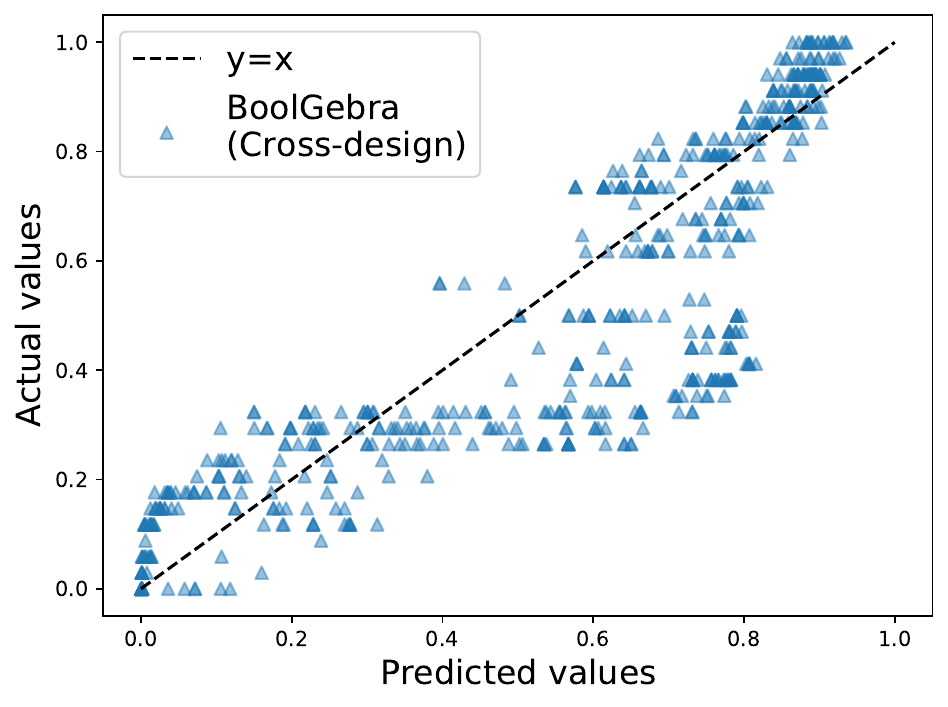}
  \caption{Testing: b12, Training: c2670}
  \label{fig:b12_c2670}
\end{subfigure}%
\hfill
\begin{subfigure}{0.32\textwidth}
  \centering
  \includegraphics[width=\linewidth]{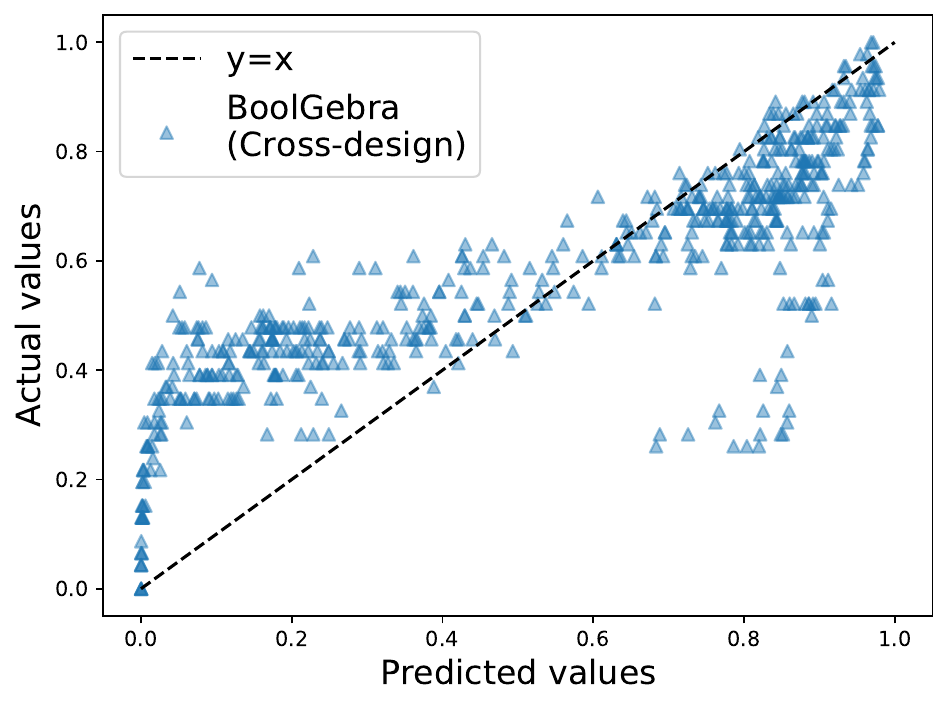}
  \caption{Testing: b11, Training: c2670}
  \label{fig:b11_c2670}
\end{subfigure}
\hfill
\begin{subfigure}{0.32\textwidth}
  \centering
  \includegraphics[width=\linewidth]{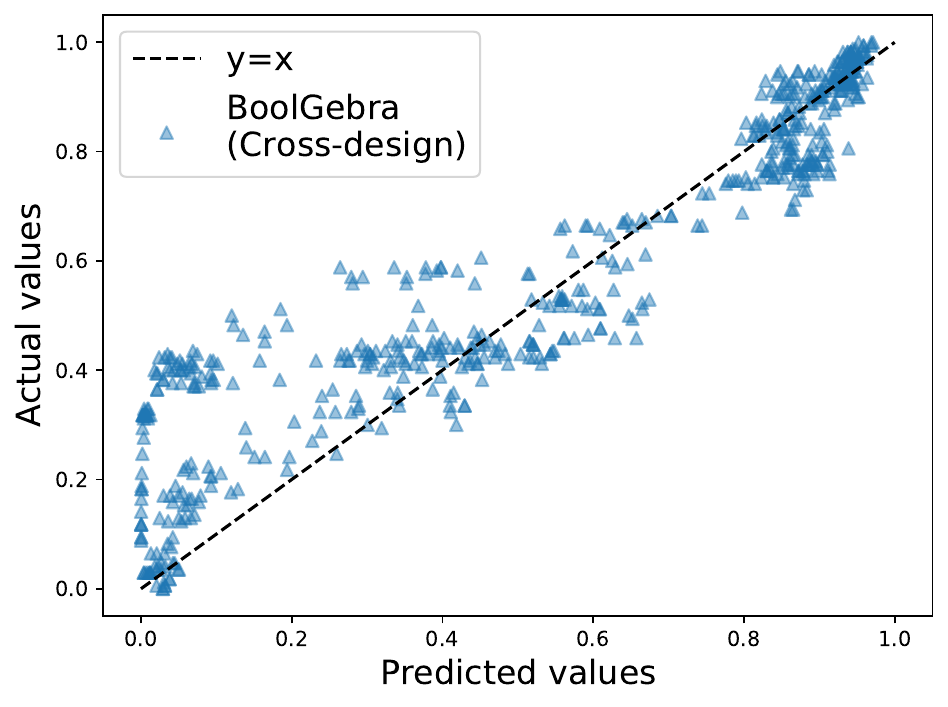}
  \caption{Testing: c5315, Training: c2670}
  \label{fig:c5315_c2670}
\end{subfigure}
\hfill

\begin{subfigure}{0.32\textwidth}
  \centering
  \includegraphics[width=\linewidth]{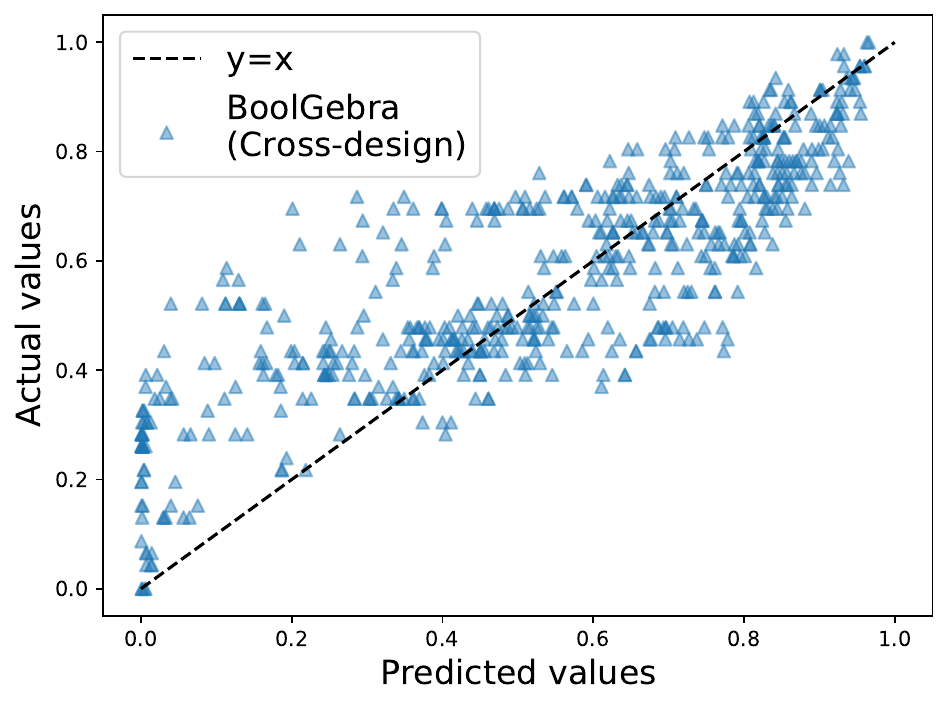}
  \caption{Testing: b11, Training: c5315}
  \label{fig:b11_c5315}
\end{subfigure}%
\hfill
\begin{subfigure}{0.32\textwidth}
  \centering
  \includegraphics[width=\linewidth]{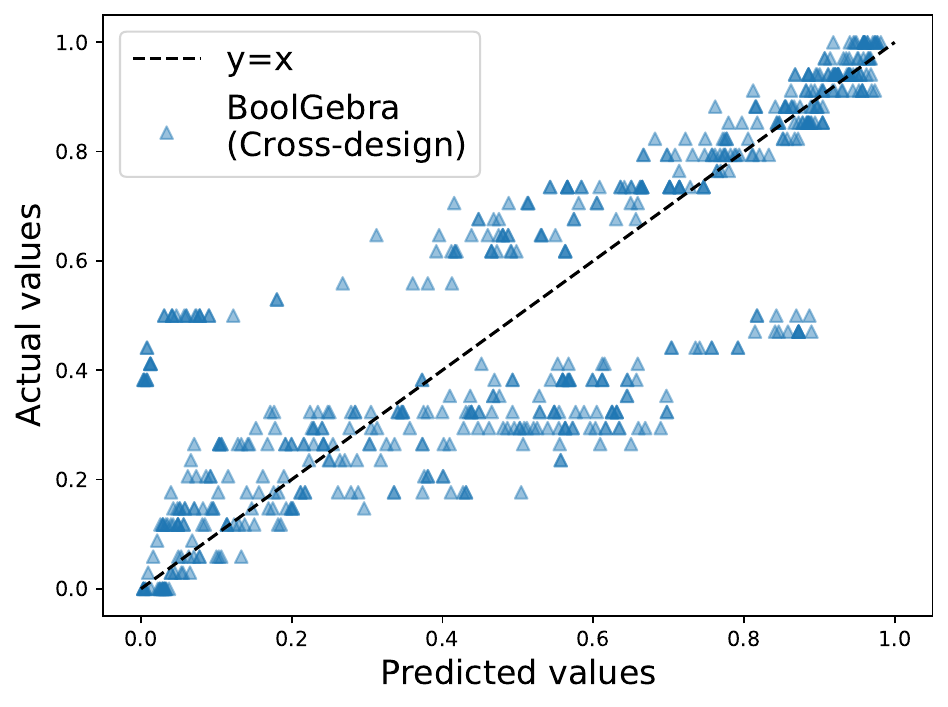}
  \caption{Testing: b12, Training: c5315}
  \label{fig:b12_c5315}
\end{subfigure}
\hfill
\begin{subfigure}{0.32\textwidth}
  \centering
  \includegraphics[width=\linewidth]{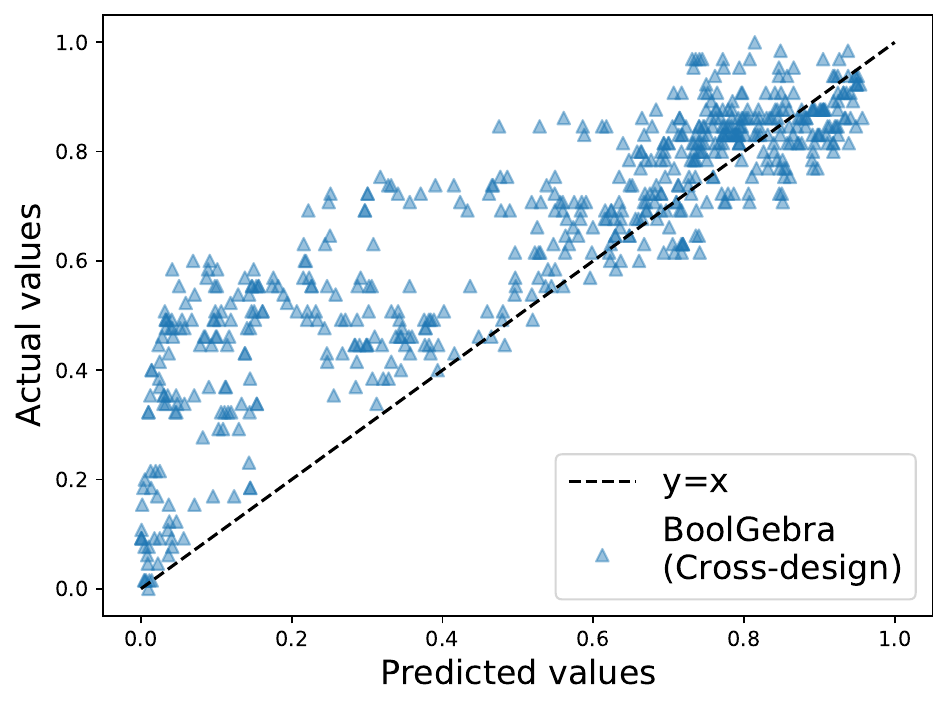}
  \caption{Testing: c2670, Training: c5315}
  \label{fig:c2670_c5315}
\end{subfigure}

\caption{Cross-design inference evaluation for predicting AIG minimization performance for given manipulation decisions. Each sub-figure presents the inference results versus the ground truth, where the model is trained on one design (Training) and tested on unseen design (Testing), denoted in the sub-captions.}
\label{fig:cross_design_predictions}
\vspace{-3mm}

\end{figure*}

In this section, we present the experimental results of the proposed GNN-based AIG prediction model w.r.t its training performance on different designs, its design specific inference for each training design, and its cross design inference for different designs. Furthermore, we provide the comparison between the model selected optimization samples with the SOTA stand-alone optimizations to demonstrate the potential of the model in shrinking the search space and locating better optimization samples w.r.t various designs. 
{All the experiments are conducted on Intel Xeon Gold 6230 20x CPU with RTX 3090 Ti GPU.}

\subsection{Design-Specific Evaluation}
\label{sec:specific_eval}

The proposed model in Section \ref{sec: predict_model} is trained with the dataset with $600$ priority guided random samples w.r.t each design. {The training batch size is 100, using Adam \cite{kingma2014adam} optimization algorithm with learning rate $8\cdot10^{-7}$. The decay rate is 0.5 for every 100 epochs.}
Specifically, testing loss curves over $1500$ training epochs with dataset of design \texttt{b07}, \texttt{b08}, \texttt{b09}, \texttt{b10}, \texttt{b12}, \texttt{c2670}, \texttt{c5315} from ISCAS85 and ITC/ISCAS99 benchmarks, are shown in Figure \ref{fig:training_loss}. The model can coverage efficiently with all of the datasets, while the model trained with \texttt{b12} and \texttt{c5315} can result in better prediction performance with the smaller loss value.  

Additionally, we visualize the prediction result versus the ground truth labels w.r.t each design in Figure \ref{fig:specific_eval}, where the predicted values are the inference results of the training dataset with the same design specific trained model. For design \texttt{b07} and design \texttt{b10} as shown in Figure \ref{fig:spec_b07} and \ref{fig:spec_b10}, the ground truth labels are discrete due to its small node size (i.e., $366$ and $180$, respectively),  and less graph structures for optimizations, resulting in worse model training performance. For design \texttt{c2670} in Figure \ref{fig:spec_c2670}, the prediction performs better at larger ground truth labels, while for better cases with smaller ground truth labels, the model can produce over-positive predicted results than actual labels as the training dataset contains much less data samples with better optimization, i.e., smaller ground truth labels (as shown in the sampling distribution in Figure \ref{fig:sample}) than the others, thus less accurate at the inference of better optimization samples. However, note that we do not expect to acquire exact optimization results from the model, we can still use the model to locate the most promising optimization samples. The model trained with design \texttt{b11}, \texttt{b12} and \texttt{c5315} show impressive correlation between the actual values and the predicted values.


\subsection{Cross-Design Evaluation}
\label{sec:cross_eval}
Furthermore, we conduct the cross-design evaluations with combinations of different training designs and testing designs. The results are shown in Figure \ref{fig:cross_design_predictions}, including 9 combinations of training designs \texttt{b11}, \texttt{c2670}, and \texttt{c5315}, testing designs \texttt{b11}, \texttt{b12}, \texttt{c2670}, and \texttt{c5315}. The correlation trend in cross design inference is similar to the trend in design specific inference. Specifically, the model trained with design \texttt{b11} shows better generalizability to other designs than other training designs. As shown in Figure \ref{fig:b12_b11} -- \ref{fig:c5315_b11}, the cross design evaluations show clean clustering trend similar to the design specific evaluations, which indicates the generalization capability of the proposed GNN-based prediction model.

\subsection{Improvements over SOTA stand-alone optimizations}
\label{sec:improve}

{ 
This section presents a comparative analysis between the exact node optimization results and the original And-Inverter Graph (AIG) node size. The top 10 optimized samples, as predicted by the model, were exclusively trained on a single design known as 'b11'. The corresponding outcomes are documented in Table 1. Specifically, we have included both the average performance (BG-Mean) of the predicted top 10 samples and the best result (BG-Best).

\textbf{It should be noted that since the training was based on the 'b11' design}, sample selections for other designs were inferred through cross-design considerations. When contrasted with state-of-the-art (SOTA) stand-alone optimizations like 'rw,' 'rs,' and 'rf,' the selected samples outperformed with a demonstrable improvement in AIG optimization. Specifically, enhancements were measured at 3.6\%, 5.3\%, and 5.5\% respectively, underscoring the potential of the BoolGebra flow. We want to point out that an average 3--5\% logic minimization gain w.r.t SOTA logic synthesis is fundamentally significant.
}


\begin{table}[htbp]
\centering
\caption{Boolean minimization evaluations compared to state-of-the-art (SOTA) synthesis methodologies \texttt{rewrite}, \texttt{resub}, and \texttt{refactor} in ABC\cite{mishchenko2007abc} framework. The results are presented in optimized AIG sizes (\%) over the original AIG sizes. \textbf{Impr.(\%)} refers improvements gained with BoolGebra-Best over the three SOTA methods. \textbf{BG} refers to BoolGebra.}
\begin{tabular}{|l|l|l|l|l|l|}
\hline
\multicolumn{1}{|l|}{Designs} & \multicolumn{1}{l|}{\texttt{rewrite}} & \multicolumn{1}{l|}{\texttt{resub}} & \multicolumn{1}{l|}{\texttt{refactor}} & \multicolumn{1}{c|}{\begin{tabular}[c]{@{}c@{}}BG\\ (Mean)\end{tabular}} & \multicolumn{1}{c|}{\begin{tabular}[c]{@{}c@{}}BG\\ (Best)\end{tabular}} \\ \hline
\hline
b07 & 0.981 & 0.975 & 0.959 & 0.940 & 0.934 \\
b08 & 0.935 & 0.923 & 0.987 & 0.917 & 0.910 \\
b09 & 0.978 & 0.971 & 0.993 & 0.956 & 0.956 \\
b10 & 0.978 & 0.950 & 0.978 & 0.937 & 0.933 \\
b11 & 0.895 & 0.897 & 0.881 & 0.834 & 0.828 \\
b12 & 0.968 & 0.964 & 0.988 & 0.950 & 0.950 \\
c2670 & 0.824 & 0.895 & 0.862 & 0.798 & 0.794 \\
c5315 & 0.836 & 0.958 & 0.893 & 0.804 & 0.801 \\
\hline
Avg & 0.925 & 0.942 & 0.943 & 0.892 & 0.888 \\
\textbf{Impr.} & \textbf{3.6\%} & \textbf{5.3\%} & \textbf{5.5\%} & - & -\\
\hline
\end{tabular}
\label{tab:improve}
\end{table}


\section{Conclusion}
\label{sec:conclusion}

This work proposes a novel GNN-based AIG algorithmic space exploration flow BoolGebra to identify novel Boolean manipulation decisions for the foundation of logic minimization. By introducing random sampling techniques, optimization and functional aware graph feature embedding, and the GNN model, BoolGebra can effectively identify top optimization candidates in the large random sample batch. 
Our model shows its potential in producing the promising optimization samples not only for design specific inferences but also for cross-design inferences. Our cross-design evaluation (trained with only one design) outperforms the three SOTA AIG optimization techniques with 3.6\%, 5.3\%, and 5.5\% improvements. Our future work will focus on improving BoolGebra to identify novel Boolean manipulation solutions for technology-dependent stages such as mapping and placement.

\noindent
\textbf{Acknowledgement:} This work is supported by National Science Foundation NSF-2047176, NSF-2019336, NSF-2008144, and NSF-2229562
awards.


\newpage

\bibliographystyle{IEEEtran}
\bibliography{ref}

\end{document}